\documentclass[
a4paper,10pt]{article}
\expandafter\let\csname equation*\endcsname\relax
\expandafter\let\csname endequation*\endcsname\relax 
\usepackage{graphicx}
\usepackage[latin1]{inputenc}
\usepackage{epsfig}
\usepackage{amsmath}
\usepackage{amsfonts}
\usepackage{float}
\usepackage{amssymb}
\usepackage{color}
\usepackage[colorlinks,citecolor=blue,urlcolor=blue,linkcolor=blue]{hyperref}
\usepackage{array,multirow,makecell}
\begin{document}

\begin{center}
	{\LARGE\bf  Alleviating the $\bar B \to D \tau \nu_\tau$ and $\bar B \to D^* \tau \nu_\tau$
		\vskip 0.3cm
		puzzle in the MSSM  }
	\vskip 1cm
	$
	\textrm{\Large Dris Boubaa$^{ 1,2,3}$ , Shaaban Khalil$^{3}$ and Stefano Moretti$^{1}$ }
	$
	\vskip 0.7cm
	{\it $^1$ School of Physics and Astronomy, University of Southampton, Highfield, Southampton SO17 1BJ, UK\\
	$^2$
	D\'epartement de Physique, Facult\'e des Sciences Exactes et Informatique,
	Universit\'e Hassiba Benbouali de Chlef, P${\hat{\rm o}}$le Universitaire d'Ouled Fares, 02180, Chlef, Alg\'erie\\
	$^3$
	Center for Fundamental Physics, Zewail City of
	Science and Technology, Sheikh Zayed,12588, Giza, Egypt	} \\
	\vskip 0.4cm
	{E-mail: {d.boubaa@soton.ac.uk\\
				skhalil@zewailcity.edu.eg\\
				s.moretti@soton.ac.uk }}
\end{center}

\begin{abstract}
We show that Supersymmetric effects driven by penguin contributions to the $%
b \to c \tau \nu_\tau$ transition are able to account simultaneously for a sizeable increase of both
branching ratios of $\bar{B}\to D \tau\bar{\nu}_{\tau}$ and $\bar{B}%
\to D^{*} \tau\bar{\nu}_{\tau}$ with respect to the Standard Model predictions, thereby approaching their experimentally measured values. We emphasise that a light chargino and neutralino, with masses less than 300 GeV, in addition to a large stau/snueutrino mass and a large $\tan\beta$, are essential for enhancing the effect of the lepton penguin $\tau \nu_\tau W^{\pm}$, which is responsible for the improved theoretical predictions with respect to current data.
\end{abstract}


\section{Introduction}\label{sec1}

Rare $B$-decays provide a good opportunity for probing New Physics (NP)
Beyond the Standard Model (BSM). In fact, experimental studies of flavour at
(Super) $B$-factories (BaBar and Belle) and LHCb are complementary to the
direct search for NP at the Large Hadron Collider (LHC). The origin of
flavour and Charge-Parity (CP) violation is one of the most profound open questions in
particle physics. Most extensions of the SM, wherein the latter is
embedded as a low energy effective theory, include new sources of flavour
and CP violation. Supersymmetry (SUSY) is one  promising candidate for
BSM physics which has these characteristics, particularly if the soft SUSY-breaking terms are
non-universal.

It has been recently reported a deviation from the SM expectations in the
ratios
\begin{equation}
{R}(D)=\frac{{\rm BR}(\bar{B}\rightarrow D\tau\bar{\nu}_{\tau})}{{\rm BR}(\bar{%
B}\rightarrow Dl\bar{\nu}_{l})}, ~~ {R}(D^{\ast})=\frac{{\rm BR}(%
\bar{B}\rightarrow D^{\ast}\tau\bar{\nu}_{\tau})}{{\rm BR}(\bar{B}\rightarrow
D^{\ast}l\bar{\nu}_{l})},
\end{equation}
where $l$ refers to either electron or muon. On the one hand, between 2015 and 2017, the Belle
collaboration  \cite{Huschle:2015rga,Sato:2016svk,Abdesselam:2016cgx,Hirose:2016wfn} has reported the following results:
\begin{align}
{R}(D)^{\rm Belle} &=0.375\pm0.064,\\ 
{R}(D^{\ast })^{\rm Belle}_{\rm ave}&=0.288\pm0.019.
\end{align}
On the other hand, the BaBar collaboration found that \cite{Lees:2013uzd}
\begin{align}
{R}(D)^{\rm BaBar} &=0.440\pm0.072,\\
{R}(D^{\ast })^{\rm BaBar} &=0.332\pm0.030.
\end{align}
In addition, the LHCb collaboration has announced the following value for ${R}(D^{\ast })$ \cite{Aaij:2015yra}:
 \begin{equation}
{R}(D^*)^{\rm LHCb} = 0.336 \pm 0.027 \pm 0.030.
 \end{equation}
Therefore, the overall combined average is given by \cite{Amhis:2016xyh}: 
\begin{align}
{R}(D)&=0.403\pm0.040,\\
{R}(D^{\ast})&=0.310\pm0.015,
\end{align}
which deviate by $\sim1.7\sigma$ for ${R}(D)$ and $\sim3.9\sigma$ for ${R}(D^{\ast})$ from the SM expectations that are given 
by \cite{Tanaka:2012nw,Sakaki:2013bfa,Sakaki:2014sea}
\begin{align}
{R}(D)^{\rm SM}&=0.305\pm0.012,\\
{R}(D^{\ast})^{\rm SM}&=0.252\pm0.004.
\end{align}
These deviations, if confirmed, could be important hints
for NP, especially because the SM results for ${R}(D)$ and ${R}(D^{*})$ are
essentially independent of the parameterisation of the hadronic
matrix elements.

As the semileptonic decay $b\rightarrow c\tau\nu_\tau$ takes
place in the SM at tree level, it is naively expected that any BSM
contribution would be subdominant, even those embedding a charged
Higgs boson entering at tree-level, since $M_{H^\pm}\ge M_{W^\pm}$.
Indeed, it is notoriously  challenging to account for large deviations from
the SM rates. This has been shown explicitly in some SM extensions
\cite{Fajfer:2012vx,Crivellin:2012ye,Crivellin:2013wna,Celis:2012dk,Celis:2016azn}. In particular,
it was emphasised that  in 2-Higgs Doublet Models (2HDMs) the
above experimental results for ${R}(D)$ and ${R}%
(D^{\ast})$ cannot be simultaneously explained.

In this article we argue that SUSY contributions, as described in
the Minimal Supersymmetric Standard Model  (MSSM) with non-universal
soft SUSY-breaking terms, might help to explain the discrepancy between the
experimental results for $\bar{B}\rightarrow D\tau\nu_{\tau}$ and
$\bar{B}\rightarrow D^{\ast}\tau\nu_{\tau}$ and the corresponding SM
expectations. For the first time in literature, to our knowledge, we
consider here all contributions up to Next-to-Leading Order (NLO)
within the MSSM: tree-level ones due to charged gauge boson and
Higgs exchange as well as one-loop ones due to bubbles,
triangles (penguins) and boxes onset by the exchanges of
2HDM states (i.e., $\gamma$, $Z$, $W^\pm$, $H^\pm$, $h^0$, $H^0$ and $A^0$)
alongside the SUSY ones due to gauginos (charginos and neutralinos)  and sfermions
(squarks, sleptons and sneutrinos). Our results ameliorate the situation with respect to the 
aforementioned data, yet even higher orders may be required to achieve full consistency.

The plan of the paper is as follows. In the next section we describe the calculation in some detail in terms of helicity amplitudes and corresponding observables. In Sect.~\ref{sec3} we introduce the Wilson coefficients needed for the calculation. Then we describe the experimental constraints enforced and illustrate our numerical analysis. We finally  conclude in Sect.~\ref{sec6}.

\section{Model Independent Contributions to $R(D)$ and $R({D^{\ast}})$ }\label{sec2}

The effective Hamiltonian for $b\rightarrow cl\bar{\nu_{l}}$ is given by
\begin{eqnarray}
{\mathcal{H}}_{\mathrm{eff}} &=& \frac{4G_{F}V_{cb}}{\sqrt{2}}\Big{[}%
(1+g_{VL})[\bar{c}\gamma_{\mu}P_{L}b][\bar{l}\gamma_{\mu}P_{L}\nu_{l}]
+ g_{VR} [\bar{c}\gamma_{\mu}P_{R}b][\bar{l}\gamma_{\mu}P_{L}%
\nu_{l}]\nonumber\\
& +&g_{SL} [\bar{c}P_{L}b][\bar{l}P_{L}\nu_{l}]  
 + g_{SR} [\bar{c}P_{R}b][\bar{l}P_{L}\nu_{l}]+g_{T} [\bar{c}\sigma^{\mu\nu_\tau
}P_{L}b][\bar{l}\sigma_{\mu\nu}P_{L}\nu_{l}]\Big{]},~~~~
\end{eqnarray}
where $G_{F}$ is the Fermi coupling constant,  $V_{cb}$ is the Cabibbo-Kobayashi-Maskawa (CKM) matrix element between charm and bottom quarks while $P_{L/R}=(1-/+\gamma_{5})/2$.
Further, $g_i$ is defined in terms of the Wilson coefficients  (see \cite{Bhattacharya:2015ida} for prospects of extracting
these using optimal observables) $C_i$ as %
\begin{equation}
g_{i}=\frac{C_{i}^{\rm SUSY}}{C^{\mathrm{SM}}},~~ ~~ i\equiv VL, VR, SL, SR, T
\end{equation}
with $C^{\mathrm{SM}}=\frac{4G_{F}V_{cb}}{\sqrt{2}}$. Therefore, the full amplitude $\bar{B}\rightarrow D l\bar{\nu}_{l}$ takes the form
%
\begin{equation}
\mathcal{M}=\mathcal{M}_{\rm SM}^{\lambda_{D^{(\ast)}}%
	,\lambda_{l}}+\mathcal{M}_{S(L,R)}^{\lambda_{D^{(\ast)}},\lambda_{l}}%
+\mathcal{M}_{V(L,R)}^{\lambda_{D^{(\ast)}},\lambda_{l}}+\mathcal{M}_{T}%
^{\lambda_{D^{(\ast)}},\lambda_{l}},
\end{equation}
where $\lambda_{l}$ is the helicity of the lepton $l$. The $D^{(\ast)}$-meson is taken to be either a spin-0
$D$-meson, with $\lambda_{D}=0$, or a spin-1 $D^{\ast}$-meson, with
$\lambda_{D^{\ast}}=\pm,0$.

Furthermore, one can define both obsevables $R(D)$ and
$R(D^*)$ as follows:
\begin{equation}
R(D)=\frac{\Gamma(\bar{B}\rightarrow D\tau\nu_{\tau})}{\Gamma(\bar{B}\rightarrow Dl\nu_{l})},~~~R(D^{\ast})=\frac{\Gamma(\bar{B}\rightarrow D^{\ast}\tau\nu_{\tau})}{\Gamma(\bar{B}\rightarrow D^{\ast}l\nu_{l})}.
\end{equation}
\begin{figure}[!ht]
	\centering
	\includegraphics[height=5cm,width=7.cm]{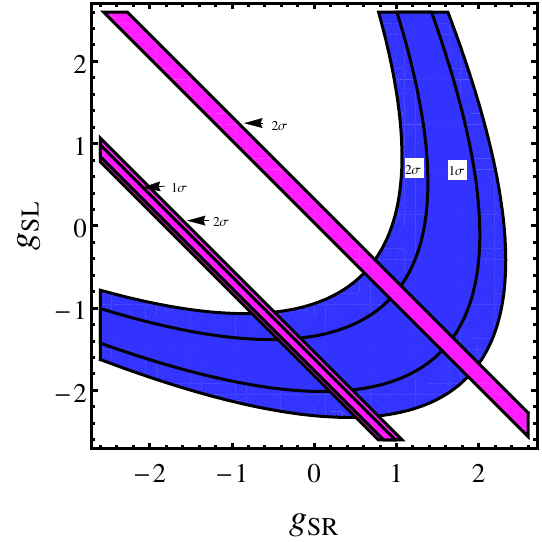}~~~ \includegraphics[height=5 cm,width=7.cm]{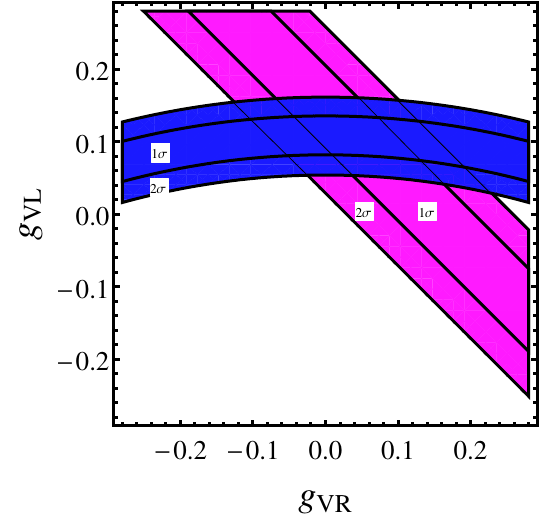}	
	\vspace{-0.1cm}
	\caption{The allowed regions in the $(g_{SL}, g_{SR})$  (left) and  $(g_{VL}, g_{VR})$  (right) planes by the $1\protect\sigma$ and $2\protect\sigma$ experimental results on ${R}(D)$ (magenta) and ${R}(D^*)$ (blue) of the combined average.}
	\label{fig1}
\end{figure}
Using the explicit formulae of the hadronic and leptonic amplitudes in Refs.~\cite{Tanaka:2012nw,Sakaki:2013bfa,Hagiwara:1989cu,Hagiwara:1989gza,Datta:2012qk,Duraisamy:2013kcw} (when the $l$ contribution is assumed to be described by the SM\footnote{This assumption is made here only for convenience, so as to write model independent analytical formulae. In the next sections though, SUSY contributions are analysed for all processes: $\bar{B}\rightarrow D\tau\nu_{\tau}$ and $\bar{B}\rightarrow Dl\nu_{l}$, i.e.,
$$R(D) = \frac{{\rm BR}(B \to D \tau \nu)^{\rm SM+SUSY}}{{\rm BR}(B \to D l \nu)^{\rm SM+SUSY}}.$$ 
In presence of experimental constraints on BR$(\bar{B}\rightarrow Dl\nu_{l})$, which are in fact quite consistent with the SM results, i.e.,
BR$(B \to D l \nu)^{\rm SM+SUSY}$ is within the experimental range of the measured BR$(B \to D l \nu)$
 (similar arguments hold for the $B^*$ rates).}) and upon fixing the SM parameters as well as 
the form factors involved in the definition of the matrix elements
to their central values as in Ref.~\cite{Lees:2013uzd}, we can cast
the explicit dependence of  ${R}(D)$ and
${R}(D^*)$ upon the Wilson coefficients in the MSSM as follows:
\begin{align}
R(D)&  ={R}(D)^{\mathrm{SM}}\Big[0.981|g_{SR}+g_{SL}|^{2}%
+|1+g_{VL} +g_{VR}|^{2} +0.811 |g_{T}|^{2}\nonumber\\ &+1.465 \operatorname{Re}[(1+g_{VL}+g_{VR})
\times(g_{SR}+g_{SL})^{\ast}]
+1.074\operatorname{Re}[(1+g_{VL}+g_{VR})g_{T}^{\ast}]\Big],\\
R({D^{\ast}}) &  ={R}(D^{\ast})^{\mathrm{SM}}\Big[0.025 |g_{SR}%
-g_{SL}|^{2}+|1+g_{VL}|^{2}+|g_{VR}|^{2}
+16.739|g_{T}|^{2}\nonumber\\
&+0.094\operatorname{Re}[(1+g_{VL}+g_{VR})
\times(g_{SR}-g_{SL})^{\ast}]
+6.513\operatorname{Re}[g_{VR}g_{T}^{\tau\ast}]\nonumber\\
&-4.457\operatorname{Re}[(1+g_{VL})g_{T}^{\ast}] -1.748\operatorname{Re}%
[(1+g_{VL})g_{VR}^{\ast}]\Big].
\end{align}
Thus, in case of a dominant scalar contribution (and negligible vector and
tensor ones), it is clear that ${R}(D^*)$ cannot be significantly larger than
the SM expectation, due to the smallness of the  coefficient of  this contribution, unless $\vert g_{SR} - g_{SL}\vert $ is much larger than 1 ({i.e.}, $C_S^{\mathrm{\rm SUSY}} > C^{\mathrm{SM}}$), which is not possible. Recall that $g_{SR}$ is larger than $g_{SL}$
and  receives a contribution at the tree-level via charged Higgs boson ($H^\pm$) exchange that yields
\begin{equation}
g_{SR}=\frac{-m_{b}m_{\tau}\tan{\beta}^{2}}{M_{H^{\pm}}^{2}}.
\end{equation}

This conclusion is confirmed in Fig.~\ref{fig1}, where we display the regions in the ($g_{SL}, g_{SR}$) plane that can accommodate the experimental
results of ${R}(D)$ and ${R}({D^*})$ within $1 \sigma$ and $2 \sigma$  Confidence Level (CL) for, e.g., Belle, the experiment with predictions closer to the SM.
From this figure, it is clear that the scalar contribution alone cannot account for both ${R}(D)$ and ${R}({D^*})$
simultaneously. In order to get ${R}(D)$ and ${R}({D^*})$ within $2\sigma$ of the  
aforementioned average results from the various experiments, $(g_{SL}, g_{SR})$ should lie between
$[-2.32, -0.77]$ and $[-0.39, 2.03]$, respectively. In these conditions, either $g_{SL}$ or $g_{SR}$ is larger than 1, which is not possible.

In case of a dominant vector contribution, as shown from the allowed regions of  ($g_{VL}, g_{VR}$) in Fig.~\ref{fig1}, one gets ${R}(D)$ and ${R}(D^*) $ inside the $2 \sigma$ region of the averages if $(g_{VL}, g_{VR})$ varies between $(0.05,0.02)$ and $(0.15,0.10)$, respectively.
Furthermore, it is remarkable that, unlike the scalar contribution, a small vector contribution, $g_{VL}$ $\sim$ $ {\cal O}(0.1)$ and $g_{VR} \sim {\cal O}(0.01)$, can induce significant enhancement for both ${R}(D) $ and ${R}(D^*) $: e.g.,  ${R}(D) \sim 0.336$ and ${R}(D^*) \sim 0.277$ if $g_{VL} \sim 0.05$ and $g_{VR} \sim 0$, which, as we will see, are quite plausible values in the MSSM. Finally, the tensor contribution, which is typically quite small, may affect only ${R}(D^*)$.

%
\section{SUSY Contributions to $b\to c\tau\nu_{\tau}$ }\label{sec3}

\begin{figure}[!t]
	\centering \epsfig{file=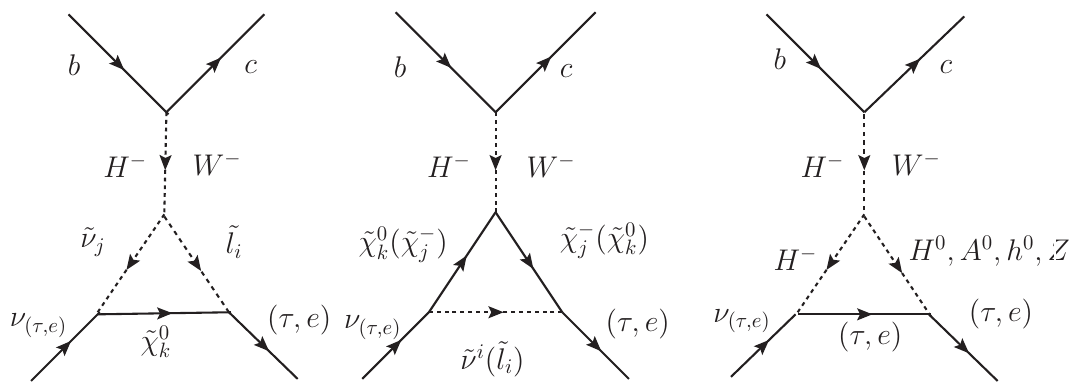,height=3.5cm,width=10cm,angle=0}
	\caption{Triangle diagrams (penguins) contributing to, {e.g.}, $b
		\to c \protect(\tau,e) \protect\nu_{(\tau,e)}$ affecting the leptonic vertex.
	}
	\label{fig2}
\end{figure}

The SUSY contributions to $g_{VL}$ are generated from the penguin corrections to the vertex $W^\pm l \nu_l $ $(l=e,\mu,\tau)$ through the exchange of charginos and neutralinos alongside sleptons and sneutrinos, respectively, as displayed in Fig.~\ref{fig2}. Our calculation  is based on  FlavorKit \cite{Porod:2014xia}, SARAH \cite{Staub:2013tta} and SPheno
\cite{Porod:2011nf}, although the dominant penguin corrections were also derived analytically.
{Renormalisation is performed at one loop using the $\overline{\rm DR}$ scheme (following SARAH and SPheno) including the full momentum dependence for any SUSY and Higgs state. As a cross-check of the implementation, we have explicitly verified that, while our loop integrals for the two and three point functions depend upon the renormalisation scale, such a dependence drops out in the computation of physical observables. In fact, it can be extracted from our equations that the loop corrections scale with $v^{2}/M_{\rm SUSY}^{2}$, so that, in the limit of very large $M_{\rm SUSY}$,  the loop effects go to zero, hence ${R}(D)$ and ${R}(D^{\ast})$ approach their SM values.
	In order to have sizable loop functions, we will enforce on our scans  the condition $m_{\chi^{0}_1}\approx m_{\chi^{-}_1}\lesssim 500~{\rm GeV}$}.

%


{Let us now try to decode our results, by concentrating on the
	Wilson coefficient $C_{VL}$, which sees contributions induced by the penguin topologies in Fig. \ref{fig2}.
	Firstly, we can confirm that the graph with neutral Higgs bosons is small while the other two
	are roughly comparable.} Thus, the emerging $C_{VL}^{\mathrm{SUSY}}$ term is essentially

\begin{equation}\label{CVL1}
C_{VL}^{\mathrm{SUSY}}=C_{VL}^{\tilde{%
		\tau}}+C_{VL}^{\tilde{\nu}} +C_{VL}^{(A^{0},H^{0},h^0)},
\end{equation}%
where 
\begin{align}\label{CVL}
C_{VL}^{\tilde{\tau}} &=
\frac{\Gamma_{\tilde{\chi}_{j}^{-}\nu_{l_{I}}\tilde{\tau}^{\ast}_{i}}^{L}
	\Gamma_{\bar{\l}_{I}\tilde{\chi}_{k}^{0}\tilde{\tau}_{i}}^{R}\Gamma_{\bar{c
		}bW^{-}}^{L}}{
	16\pi^{2}M_{W^\pm}^{2}}\Big{[}\Gamma_{\tilde{\chi}_{j}^{+}\chi_{k}^{0}W^{-}}^{R}m_{\tilde{\chi}_{j}^{-}}m_{\tilde{\chi}_{k}^{0}}  C_{0}(m_{\tilde{\chi}_{k}^{0}}^{2},m_{\tilde{\chi}_{j}^{-}}^{2},m_{\tilde{\tau}_{i}}^{2})\nonumber\\
&-\Gamma_{\tilde{\chi}_{j}^{+}\tilde{\chi}_{k}^{0}W^{-}}^{L}(B_{0}(m_{\tilde{\chi}_{j}^{-}}
^{2},m_{\tilde{\chi}_{k}^{0}}^{2}) 
-2C_{00}(m_{\tilde{\chi}_{k}^{0}}^{2},m_{\tilde{\chi}_{j}^{-}}^{2},m_{\tilde{\tau}_{i}}^{2}) + m_{\tilde{\tau}_{i}}^{2}C_{0}(m_{\tilde{\chi}_{k}^{0}}^{2},m_{\tilde{\chi}_{j}^{-}}^{2},m_{\tilde{\tau}_{i}}^{2}))\Big{]},
\\
C_{VL}^{\tilde{\nu}} &=
\frac{\Gamma_{\nu_{l_{I}}\tilde{\chi}_{k}^{0}\tilde{\nu}_{i}^{\ast}}^{L}\Gamma_{\tilde{\chi}_{j}^{-}\bar{l_{I}}\tilde{\nu}_{i}}^{R}	
	\Gamma_{\bar{c}bW^{-}}^{L}}{
	16\pi^{2}M_{W^\pm}^{2}}\Big{[}-\Gamma_{\tilde{\chi}_{j}^{+}\chi_{k}^{0}W^{-}}^{L}m_{\tilde{\chi}_{j}^{-}}m_{\tilde{\chi}_{k}^{0}}  C_{0}(m_{\tilde{\chi}_{j}^{-}}^{2},m_{\tilde{\chi}_{k}^{0}}^{2},m_{\tilde{\nu}_{i}}^{2})\nonumber\\
&+\Gamma_{\tilde{\chi}_{j}^{+}\tilde{\chi}_{k}^{0}W^{-}}^{R}(B_{0}(m_{\tilde{\chi}_{k}^{0}}^{2},m_{\tilde{\chi}_{j}^{-}}
^{2}) 
-2C_{00}(m_{\tilde{\chi}_{j}^{-}}^{2},m_{\tilde{\chi}_{k}^{0}}^{2},m_{\tilde{\nu}_{i}}^{2}) + m_{\tilde{\tau}_{i}}^{2}C_{0}(m_{\tilde{\chi}_{j}^{-}}^{2},m_{\tilde{\chi}_{k}^{0}}^{2},m_{\tilde{\nu}_{i}}^{2}))\Big{]},
\\
C_{VL}^{A^{0}}  &=
\frac{2\Gamma_{\bar{\l} \nu_{\l} H^{-}}^{L}
	\Gamma_{\bar{\l}\l A^{0}}^{R}
	\Gamma_{A^{0}H^{+}W^{-}}
	\Gamma_{\bar{c
		}bW^{-}}^{L}}{
	16\pi^{2}M_{W^\pm}^{2}} C_{00}(m_{\l}^{2},M_{H^{-}}^{2},m_{A^{0}}^{2}).
\end{align}
The Wilson coefficients $C_{VL}^{(H^0,h^0)}$ can be obtained from $C_{VL}^{A^0}$ by exchanging $A^{0}\leftrightarrow (H^{0},h^0)$.
The corresponding couplings are given by
\begin{align}
\Gamma_{\tilde{\chi}_{j}^{-}\nu_{l_{I}}\widetilde{\tau}_{i}^{\ast}}^{L}  &= g(-Z_{L}^{iI\ast}
Z_{-}^{j1\ast}+\frac{m_{\l_{I}}}{\sqrt{2}M_{W^\pm}\cos\beta}Z_{L}^{i(I+3)\ast}Z_{-}^{j2
}), \label{chargino_coup}\\
\Gamma_{\bar{\l}_{I}\tilde{\chi}_{k}^{0}\widetilde{\tau}_{i}}^{R}  & =\frac{g}{\sqrt{2}%
}(Z_{L}^{iI\ast}(\tan\theta_{W}Z_{N}^{k1\ast}+Z_{N}^{k2\ast}) -\frac{m_{l_{I}}}{M_{W^\pm}\cos\beta}Z_{L}^{i(I+3)\ast}Z_{N}^{j3\ast}),\\
\Gamma_{\tilde{\chi}_{k}^{0}\nu_{l_{I}}\tilde{\nu}_{i}}^{L}  & =\frac{g}{\sqrt{2}}Z_{\nu}^{iI\ast}(\tan\theta_{W}Z_{N}^{k1}-Z_{N}^{k2}),\\
\Gamma_{\bar{\l}_{I}\chi_{j}^{-}\tilde{\nu}_{i}}^{R}  & =- gZ_{+}^{j1\ast}Z_{\nu}^{iI},~~\Gamma_{\overline{c}bW^{+}}^{L} =-\frac{g}{\sqrt{2}}V_{cb},\\
\Gamma_{\tilde{\chi}_{j}^{+}\tilde{\chi}_{k}^{0}W^{-}}^{L}  & =-g(Z_{-}^{j1}Z_{N}%
^{k2\ast}+\frac{1}{\sqrt{2}}Z_{-}^{j2}Z_{N}^{k3\ast}),\\
\Gamma_{\tilde{\chi}_{j}^{+}\tilde{\chi}_{k}^{0}W^{-}}^{R}  & =-g(Z_{+}^{j1}Z_{N}%
^{k2\ast}-\frac{1}{\sqrt{2}}Z_{+}^{j2}Z_{N}^{k4\ast}), \\
\Gamma_{\bar{\tau} \nu H^{-}}^{L} & = \frac{g m_{\tau}}{\sqrt{2}M_{W^\pm}\cos\beta}Z_{H^{-}}^{21},~~~
\Gamma_{A^{0}H^{+}W^{-}} = \frac{g}{2},\\
\Gamma_{\bar{\tau}\tau A^{0}}^{R} & =- \frac{1}{\sqrt{2}}\frac{g m_{\tau}}{\sqrt{2}M_{W^\pm}\cos\beta}Z_{A}^{21}, \\
\Gamma_{\bar{\tau}\tau H^{0}}^{R} & = \frac{1}{\sqrt{2}}\frac{g m_{\tau}}{\sqrt{2}M_{W^\pm}\cos\beta}Z_{H}^{21},\\
\Gamma_{H^{0}H^{+}W^{-}} & = \frac{g}{2}(Z_{H}^{22}Z_{H^{-}}^{22}-Z_{H}^{21}Z_{H^{-}}^{21}),\label{neutralino_coup}
\end{align}
\begin{figure}[!t]
	\centering
	\includegraphics[height=5cm,width=8cm]{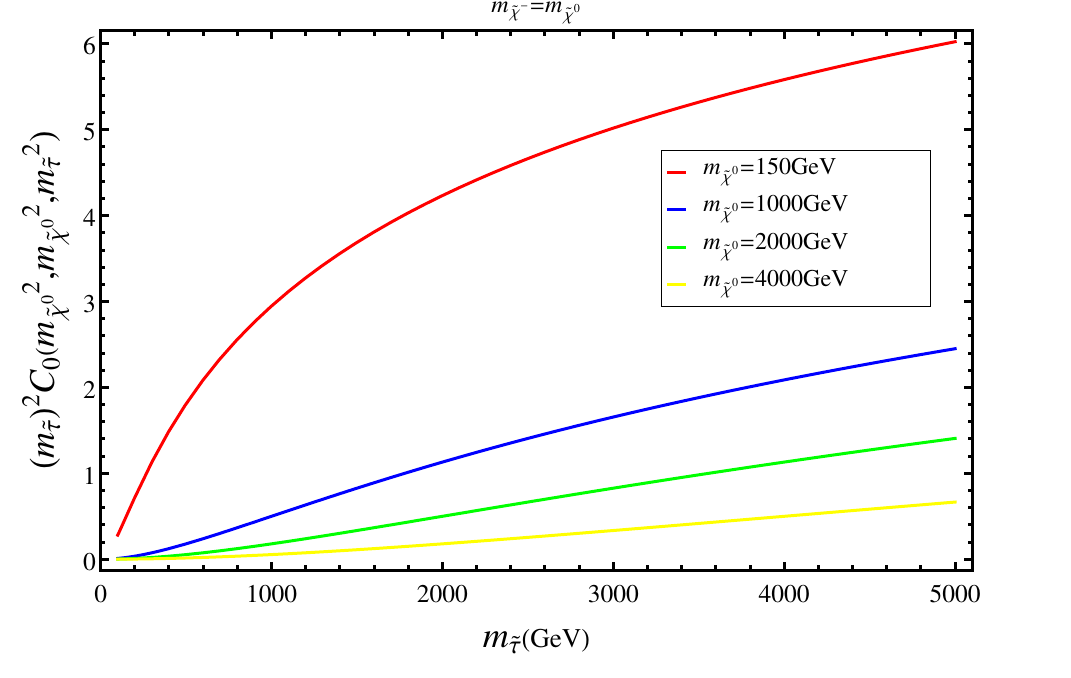}~~~
	\includegraphics[height=4.75cm,width=7.5cm]{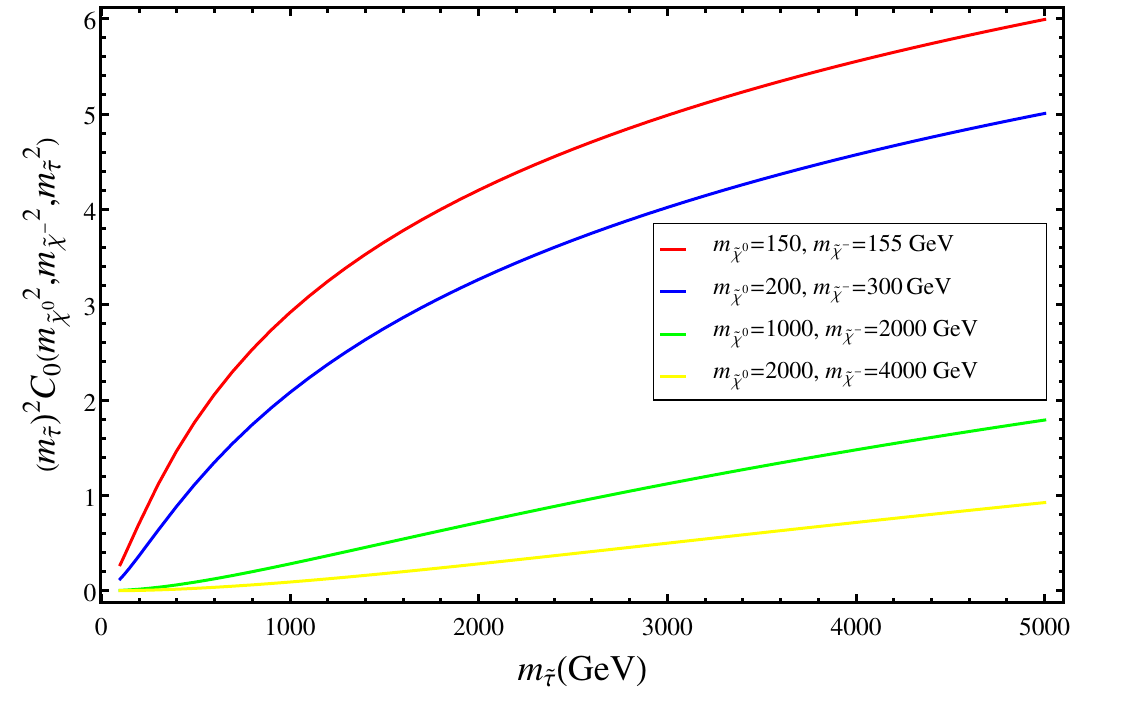}~~~
	\vspace{-0.3cm}
	\caption{Behaviour of the last term in Eq.~(\ref{CVL}), $m_{\tilde{\tau}}^2C_0(m_{\tilde{\chi}^{0}}^{2},m_{\tilde{\chi}^{-}}^{2},m_{\tilde{\tau}}^{2})$, with $m_{\tilde{\tau}}$ for degenerate (left) and non-degenerate (right) chargino/neutralino masses.}
	\label{figCo}
\end{figure}
where $Z_L$, $Z_\nu$, $Z_\pm$, $Z_N$ and $Z_{(H,A,H^{-})}$ are the diagonalising matrices for
slepton, sneutrino, chargino, neutralino and Higgs masses, respectively.
In addition, the loop functions are given by \cite{Buras:2002vd}

\begin{align}
B_{0}(x,y)& =\eta _{\varepsilon }-1+\log \frac{x}{{\tilde\mu} ^{2}}-\frac{y\log
	\frac{y}{x}}{x-y}, \\
C_{0}(x,y,z)& =\frac{1}{y-z}\Big{(}\frac{y\log \frac{y}{x}}{y-x}+\frac{z\log
	\frac{z}{x}}{x-z}\Big{)}, \\
C_{00}(x,y,z)& =\frac{1}{4}\Big{(}\eta _{\varepsilon }-\log \frac{x}{\tilde{\mu} ^{2}
}\Big{)}+\frac{3}{8}
+\frac{1}{y-z}\Big{(}\frac{y^{2}\log \frac{y}{x}}{4(x-y)}-\frac{z^{2}\log\frac{z}{x}}{4(x-z)}\Big{)},
\end{align}
with $\eta _{\varepsilon }=\frac{2}{d-4 }+\log 4\pi \gamma_{E} $, which is subtracted in the $\overline{\rm DR}$ scheme, and $\tilde \mu$  the renormalisation scale with the dimensions of mass.
Here, a few comments are in order. $(i)$ The loop function $C_{0}(m_{\tilde{\chi}_{k}^{0}}^{2},m_{\tilde{\chi}_{j}^{-}}^{2},m_{\tilde{\tau}_{i}}^{2}) \to 0$ if $m_{\tilde{\chi}_{k}^{0}}, m_{\tilde{\chi}_{j}^{-}}$ and $m_{\tilde{\tau}_{i}} \to \infty$, as expected in the SUSY decoupling limit. $(ii)$ If $m_{\tilde{\chi}_{k}^{0}}$ and $ m_{\tilde{\chi}_{j}^{-}}$ are of order ${\cal O}(100)$ GeV and $m_{\tilde{\tau}_{i}} $ is very heavy, { then $m_{\tilde{\tau}_{i}}^2 C_{0}(m_{\tilde{\chi}_{k}^{0}}^{2},m_{\tilde{\chi}_{j}^{-}}^{2},m_{\tilde{\tau}_{i}}^{2})$ does not vanish, as this is not a decoupling limit since a light fermionic SUSY spectrum is assumed}. Specifically, for
$m_{\tilde{\chi}_{k}^{0}} \simeq m_{\tilde{\chi}_{j}^{-}}$, the loop function takes the form
\begin{equation} 
C_0(m^2_{\tilde{\chi}_{i}^{0}},m^2_{\tilde{\chi}_{i}^{0}}, m^2_{\tilde{\tau}_{j}}) = \frac{1}{(m^2_{\tilde{\chi}_{i}^{0}} -m^2_{\tilde{\tau}_{j}})^2} \left[ m^2_{\tilde{\chi}_{i}^{0}} - m^2_{\tilde{\tau}_{j}} + m^2_{\tilde{\tau}_{j}} \log\left( \frac{m^2_{\tilde{\tau}_j}}{m^2_{\tilde{\chi}_{i}^{0}}}\right)\right].
\end{equation}
$(iii)$ From Eq.~(\ref{CVL}), one can see that, if $C_0(m^2_{\tilde{\chi}_{i}^{0}},m^2_{\tilde{\chi}_{i}^{0}}, m^2_{\tilde{\tau}_{j}}) \neq 0$, then the last term, proportional to $m^2_{\tilde{\tau}_{j}} C_0(m^2_{\tilde{\chi}_{i}^{0}},m^2_{\tilde{\chi}_{i}^{0}},$ $m^2_{\tilde{\tau}_{j}})$, gives the dominant effect to $C_{VL}^{\tilde{\tau}}$. $(iv)$ The typical values of the couplings $\Gamma_{\tilde{\chi}_{j}^{-}\bar{l_{I}}\tilde{\nu}_{i}}^{R}$, 	$\Gamma_{\nu_{l_{I}}\tilde{\chi}_{k}^{0}\tilde{\nu}_{i}^{\ast}}^{L}$,
$\Gamma_{\bar{c}bW^{-}}^{L}$, $\Gamma_{\tilde{\chi}_{j}^{+}\tilde{\chi}_{k}^{0}W^{-}}^{L}$ and the loop function $C_0(m^2_{\tilde{\chi}_{i}^{0}},m^2_{\tilde{\chi}_{i}^{0}}, m^2_{\tilde{\tau}_{j}}) $ at $m_{\tilde{\chi}_{i}^{0}} \sim {\cal O}(100)$ GeV and
 $ m_{\tilde{\tau}_{j}} \sim {\cal O}(1)$ TeV imply that $C_{VL}^{\tilde{\tau}} \sim
   \frac{2\times10^{-3}}{16\pi^2M_{W^\pm}^2} m^2_{\tilde{\tau}_{j}} C_0(m^2_{\tilde{\chi}_{i}^{0}},m^2_{\tilde{\chi}_{i}^{0}}, m^2_{\tilde{\tau}_{j}})$ is of order $10^{-8}~{\rm GeV}^{-2}$. Therefore, $g_{VL} = C_{VL}^{\tilde{\tau}} /C^{\rm SM}$, where $C^{\rm SM}\sim 1.38\times 10^{-6} ~{\rm GeV}^{-2}$, can be of order $0.01$.

In Fig. \ref{figCo} we show the behaviour of the last term in Eq.~(\ref{CVL}), $m^2_{\tilde{\tau}_{i}} C_0(m^2_{\tilde{\chi}_{k}^{0}},m^2_{\tilde{\chi}_{j}^{0}}, m^2_{\tilde{\tau}_{i}})$ with several examples of degenerate (left panel) and non-degenerate (right panel) chargino/neutralino masses. As it can be seen from this figure,  the largest corrections are  obtained if chargino/neutralino masses are less than 200 GeV and the stau mass is larger than 1 TeV. It is further clear that, in the light gaugino mass regime, chargino/neutralino mass degeneracy is not a pre-condition for enhancing the aforementioned loop contribution, it so happens that there can be  spectrum configurations in the scan when they are  close in mass, as allowed by experimental constraints \cite{Fuks:2017rio}.
We stress again that, even if the stau is very heavy, we are not in the decoupling limit, where SUSY effects must diminish, since charginos/neutralinos are kept quite light.

{Finally, the Wilson coefficients
$C^{\rm SUSY}(W\rightarrow\l\nu)$ and $C^{\rm SUSY}(\tau\rightarrow\nu_{\tau}\l\nu_{l})$ can be obtained from Eqs.~(\ref{CVL1}) and (\ref{CVL}) as follows
		\begin{equation}
		C^{\rm SUSY}(\tau\rightarrow\nu_{\tau}l\nu_{l})=\frac{\Gamma_{\bar{\nu}_{\tau}
				\tau W^{-}}^{L}}{\Gamma_{\bar{c
				}bW^{-}}^{L}}(C_{VL}^{\tilde{\tau}}+C_{VL}^{\tilde{\nu}}),
		\end{equation}
		\begin{equation}
		C^{\rm SUSY}(W\rightarrow\l\nu)=\frac{M_{W^\pm}^{2}}{\Gamma_{\bar{c
				}bW^{-}}^{L}}(C_{VL}^{\tilde{\tau}}+C_{VL}^{\tilde{\nu}}),
		\end{equation}
		where the loop functions  are given by the following changes:
		\begin{eqnarray}\label{loopf}
			C_{0}(0,0,0,m_{\chi_{k}^{0}}^{2},m_{\chi_{j}^{-}}^{2},m_{\tilde{\nu}_{i}}^{2}) &\rightarrow& C_{0}(m_{W}^{2},0,0,m_{\chi_{k}^{0}}^{2},m_{\chi_{j}^{-}}^{2},m_{\tilde{\nu}_{i}}^{2}),\\
			C_{00}(0,0,0,m_{\chi_{k}^{0}}^{2},m_{\chi_{j}^{-}}^{2},m_{\tilde{\nu}_{i}}^{2}) &\rightarrow& C_{00}(m_{W}^{2},0,0,m_{\chi_{k}^{0}}^{2},m_{\chi_{j}^{-}}^{2},m_{\tilde{\nu}_{i}}^{2}),\\ 
						B_{0}(0,m_{\chi_{k}^{0}}
			^{2},m_{\chi_{j}^{-}}^{2})&\rightarrow& B_{0}(m_{W}^{2},m_{\chi_{k}^{0}}
			^{2},m_{\chi_{j}^{-}}^{2}),
		\end{eqnarray}
		where the loop functions of non-vanishing $M_{W^\pm}$ can be found in  \cite{Jorge,Hollik:1988ii}.

%
\section{Experimental Constraints}\label{sec4}

{Let us now discuss   experimental limits  coming from other processes. In this regard, one should consider a possible constraint due to the direct measurement of the $W^\pm$ boson decay widths that leads to} \cite{Olive:2016xmw}
\begin{equation}\label{eq:Wen}
\Gamma(W\rightarrow\tau\nu)/\Gamma(W\rightarrow e\nu)=1.043\pm 0.024.
\end{equation}
The SM prediction for this ratio is given by $\sim 0.999267$, which is consistent with the measured value.  {{Similarly, constraints can also be obtained from}} \cite{Olive:2016xmw}
\begin{equation}\label{eq:Wmun}
\Gamma(W\rightarrow\tau\nu)/\Gamma(W\rightarrow \mu\nu)=1.07\pm 0.026,
\end{equation}
{{with which the SM is also consistent.}} The decay width of $W \to \l \nu$ with SUSY contributions can be parametrised as
\begin{equation}%
\Gamma(W\rightarrow\l\nu)=\frac{G_{F}M_{W^\pm}^{3}}{6\sqrt{2}\pi}(1-\frac{m_{l}^{2}}{M_{W^\pm}^{2}})^{2}(1+\frac{1}{2}\frac{m_{l}^{2}}{M_{W^\pm}^{2}})|1+g_{VL}^{\prime}|^{2},
\end{equation}
where $g_{VL}^{\prime}=C^{\rm SUSY}(W\rightarrow\l\nu)/C^{\rm SM}(W\rightarrow\l\nu)$ and  $C^{\rm SM}(W\rightarrow\l\nu)=g/\sqrt{2}$.
{{Another important experimental measurement connected with  lepton universality in $\tau$ decay
that should be considered here
 is of $\tau \to \nu_{\tau}l\nu_{l}$  with $l=e,\mu$, which is given by the relation\footnote{In the presence of NP, the deviations from $\tau-\mu -e$ universality can be studied via  the ratios of the branching fractions ${\rm BR}(\tau \to \nu_{\tau}e\nu_{e})/{\rm BR}(\mu \to \nu_{\mu}e\nu_{e})$, ${\rm BR}(\tau \to \nu_{\tau}\mu\nu_{\mu})/{\rm BR}(\mu \to \nu_{\mu}e\nu_{e})$, ${\rm BR}(\tau \to \nu_{\tau}\mu\nu_{\mu})/{\rm BR}(\tau \to \nu_{\tau}e\nu_{e})$, which lead to appropriate ratios $G_{\tau,e}/G_{\mu,e}$, $G_{\tau,\mu}/G_{\mu,e}$ and $G_{\tau,\mu}/G_{\tau,e}$, respectively. Here we use a different convention from those in Refs.~\cite{Chankowski:1994ds,Masiero:2008cb}, $i.e.$, we take the ratio $(g_{\mu}/g_{e})_\tau$ rather than $(G_{\tau,\mu}/G_{\tau,e})$ .} \cite{Aubert:2009qj}

\begin{equation}\label{LM}
\left(\frac{g_{\mu}}{g_{e}}\right)^{2}_\tau=\frac{BR(\tau \to \mu\nu_{\tau}\nu_{\mu})}{BR(\tau \to e\nu_{\tau}\nu_{e})}\frac{f(m_{e}^{2}/m_{\tau}^{2})}{f(m_{\mu}^{2}/m_{\tau}^{2})}.
\end{equation}
In the SM, the universal gauge interaction implies that
\begin{equation}\label{LU}
\frac{\Gamma(\tau \to \mu\nu_{\tau}\nu_{\mu})}{\Gamma(\tau \to e\nu_{\tau}\nu_{e})}=\frac{f(m_{\mu}^{2}/m_{\tau}^{2})}{f(m_{e}^{2}/m_{\tau}^{2})}=0.9726,
\end{equation}
where $f(x)=1-8x+8x^{3}-x^{4}-12x^{2}\log(x)$. The current experimental result for this ratio is $0.979\pm 0.004$ \cite{Olive:2016xmw}, which gives $\left(\frac{g_{\mu}}{g_{e}}\right)_\tau=1.0032\pm0.002$. With SUSY contributions, Eq.~(\ref{LU}) can be written as
\begin{equation}\label{LN}
\frac{\Gamma(\tau \to \mu\nu_{\tau}\nu_{\mu})}{\Gamma(\tau \to e\nu_{\tau}\nu_{e})}=0.9726\frac{|1+g_{VL}^{\mu}|^{2}}{|1+g_{VL}^{e}|^{2}},
\end{equation}
where $g_{VL}^{l}=C^{\rm SUSY}(\tau \to \nu_{\tau}l\nu_{l})/C^{\rm SM}(\tau \to \nu_{\tau}l\nu_{l})$ with $C^{\rm SM}(\tau \to \nu_{\tau}l\nu_{l})=2\sqrt{2}G_{F}$. (As we will show, this imposes stringent constraints on  SUSY contributions to $g_{VL}^l$)}.
{{Furthermore, SUSY loop effects induce a correction to the Fermi coupling via a potential
breaking of $\mu-e$ universality. In fact, using Eqs.~(\ref{LM}) and (\ref{LN}), {for $g_{VL}^l\ll 1$ one can find}
\begin{equation}\label{eq:GF}
\left(\frac{g_{\mu}}{g_{e}}\right)_\tau=\frac{|1+g_{VL}^\mu|}{|1+g_{VL}^e|}=|1+\Delta g_{VL}^{\mu,e}|,
\end{equation}
{where $\Delta g_{VL}^{\mu,e}=g_{VL}^\mu-g_{VL}^e$, so that the above experimental constraints impose that $0.0012\leq\Delta g_{VL}^{\mu,e}\leq 0.0052$.
In our work, we will enforce $g_{\mu}=g_e=g$, which  satisfies Eq.~(\ref{eq:GF}). }

{Moreover, there are other constraints that could be considered here, coming from Lepton Flavour Violating (LFV) processes such as BR$(\tau^- \to \mu^-\gamma)<4.5\times 10^{-8}$ and BR$(\tau^- \to e^-\gamma)<1.2\times 10^{-7}$ \cite{Aubert:2009ag} as well as BR$(H\to e \tau )< 1.04\%$, BR$(H\to\mu \tau ) < 1.43$ and BR$(Z\to\mu\tau) < 1.69 \times 10^{-5}$ \cite{Aad:2016blu}. {However, we will focus on the strongest one, which is indeed from the decay $\tau\to l \nu_\tau \nu_{l}$, as shown above, essentially because it carries the same one-loop corrections of the vertex $W^\pm l\nu_l$ within the process $b\to cl\nu_l$.}} 
{Furthermore, the lifetime of the $B_c$ meson may also impose important constraints on the scalar contributions, $g_{SL}$ and $g_{SR}$. However, this observable is less sensitive to the vector contribution, $g_{VL}$, which is playing an important role in enhancing $R(D)$ and $R(D^*)$ in our analysis. This has been discussed in detail in Ref.~\cite{Alonso:2016oyd}.}
In summary, in
 our scans, only points respecting all the above limits are retained. In particular, compliance with Eqs. (\ref{eq:Wen})--(\ref{eq:Wmun}) ensures that our parameter space automatically satisfies also constraints from
the ratio $\Gamma(\tau\to e\nu\nu)/\Gamma(\mu\to e\nu\nu)$. {Indeed, the SUSY contribution to $\mu\to e\nu\nu$ leads to ${\rm BR}(\mu\to e{\nu}\nu)\approx 100\%$, which is consistent with the experimental result  given in Ref. \cite{Olive:2016xmw}.}}}

{{Furthermore, the oblique Electro-Weak
(EW) parameters $S$, $T$ and $U$ \cite{Peskin:1990zt} are useful to constraint NP that enters in self-energy corrections to a gauge boson propagator, denoted by $\Pi^{}_{ij}$, which represents the transition $ij$ $(i,j=W,Z,\gamma)$, as  we have \cite{Olive:2016xmw}
\begin{equation}
\hat{\alpha}(M_Z)T=\frac{\Pi^{\rm NP}_{WW}(0)}{M^2_W}-\frac{\Pi^{\rm NP}_{ZZ}(0)}{m^2_Z},
\end{equation}
where $\hat{\alpha}(M_Z)$ is the renormalised Electro-Magnetic (EM) coupling constant at the $M_Z$ scale.
Here, we are interested in the $T$ parameter. In this respect, a related quantity known as the $\rho$ parameter is defined as \cite{Olive:2016xmw}
\begin{equation}
\rho-1=\frac{1}{1-\hat{\alpha}(M_Z)T}\simeq\hat{\alpha}(M_Z)T.
\end{equation}

In this work we take $\Delta{\rho^{\rm exp}}=\rho-1=0.0006\pm0.0009$,  which is extracted from the data on the $T$ parameter $(0.08\pm0.12)$ \cite{Olive:2016xmw}.  While in the SM $\rho\equiv\rho_0=M_{W^\pm}^2/M_Z^2\cos{\theta_W}=1$ at tree level,  in our scan we obtain $\Delta{\rho}^{\rm SUSY}\in[0.0001,0.0006]$.}}

\section{Numerical Analysis}
\label{sec5}

We now perform the numerical evaluations in the light of the results in Sects.~\ref{sec2} and \ref{sec3} in presence of  experimental constraints. Since our focus is on the penguin contributions,
let us look at the relevant loop functions entering the numerics. Let us begin with those of $W \to \l \nu$,
from the formulae given in Eq.~(\ref{loopf}) one can notice that the loop functions of the decay $W \to \l \nu$  are approximately equal to those  associated
with $b \to c \l \nu$: this is made evident in Tab.~\ref{tab1}, for the case of the MSSM benchmark of Tab.~\ref{tab2}, which is one yielding sizable corrections to both ${R}(D)$ and ${R}(D^*)$.

\begin{figure}[!h]
	\centering
	\includegraphics[height=3.5cm,width=5cm]{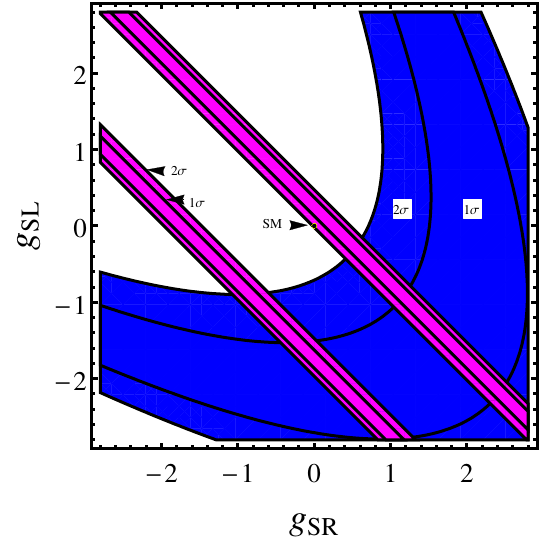}~~~
	\includegraphics[height=3.5cm,width=5cm]{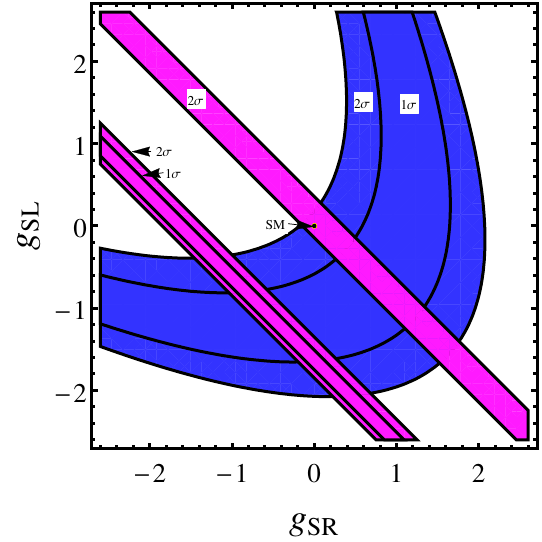}~~~
	\includegraphics[height=3.5cm,width=5cm]{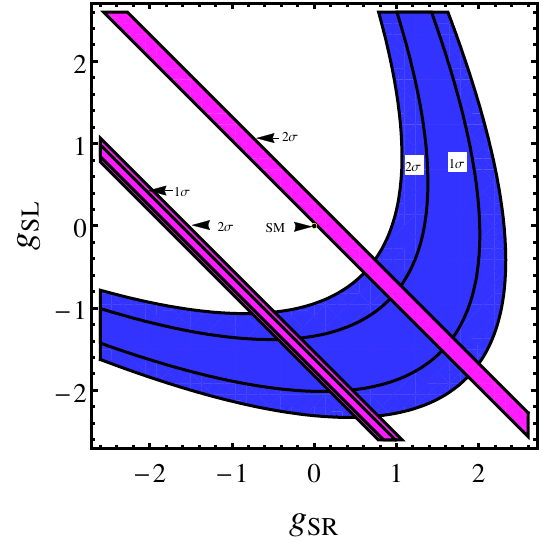}
	\vspace{-0.35cm}
	\caption{The allowed regions in the $(g_{SL}, g_{SR})$ plane by the $1\protect\sigma$ and $2\protect\sigma$ experimental results on ${R}(D)$ (magenta) and ${R}(D^*)$ (blue)
		of BaBar (left), Belle (middle) and combined average (right).  SM and complete SUSY predictions (tree, penguin and box contributions, where tree-level effect is the dominant for the scalar interactions) are also included and they coincide with the black point.  }
	\label{fig4}
\end{figure}
\begin{figure}[!h]
	\centering
	\includegraphics[height=3.4cm,width=5cm]{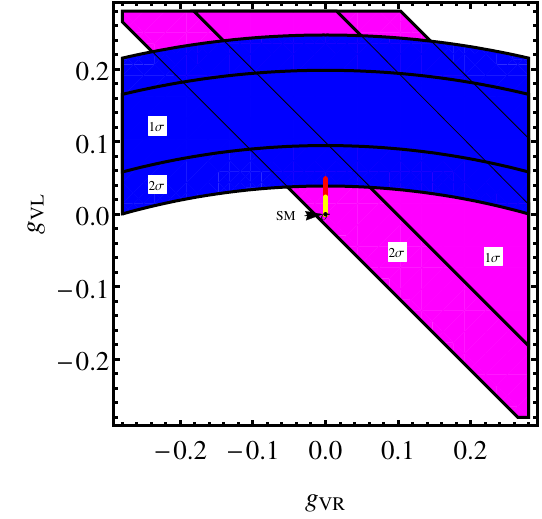}~~~
	\includegraphics[height=3.4cm,width=5cm]{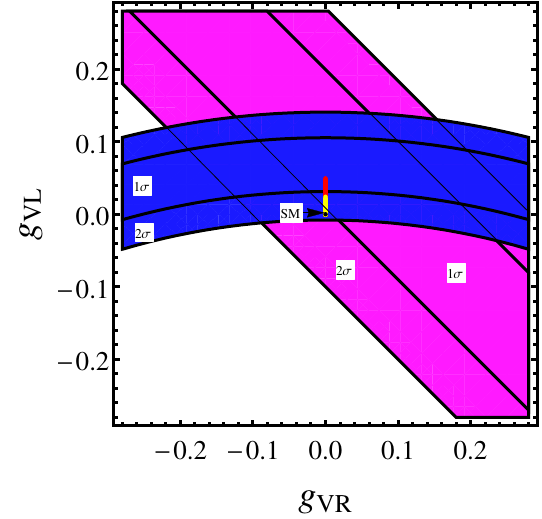}~~~
	\includegraphics[height=3.5cm,width=5cm]{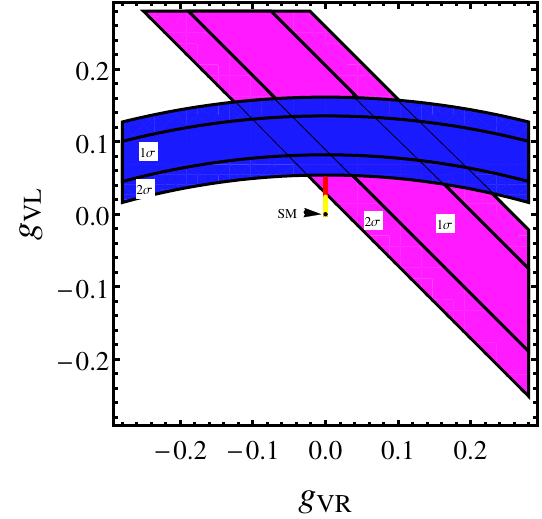}
	\vspace{-0.35cm}
	\caption{The allowed regions in the $(g_{VL}, g_{VR})$ plane by the $1\protect\sigma$ and $2\protect\sigma$ experimental results on ${R}(D)$ (magenta) and ${R}(D^*)$ (blue)
		of BaBar (left), Belle (middle) and combined average (right). SM (black point) and complete SUSY (red points) predictions (penguin and box contributions, where the penguin is the dominant one) are also included.}
	\label{fig5}
\end{figure}
\begin{figure}[!h]
	\centering
	\includegraphics[height=5cm,width=6.5cm]{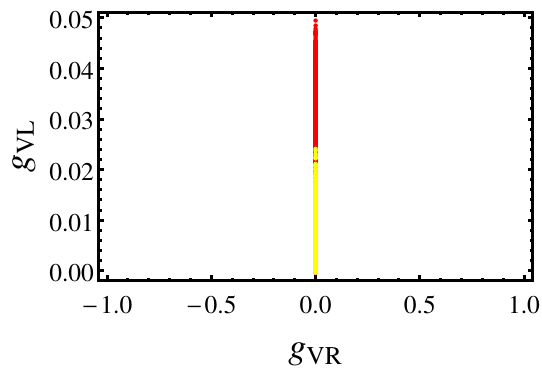}~~~ ~~~~
	\includegraphics[height=5cm,width=6.5cm]{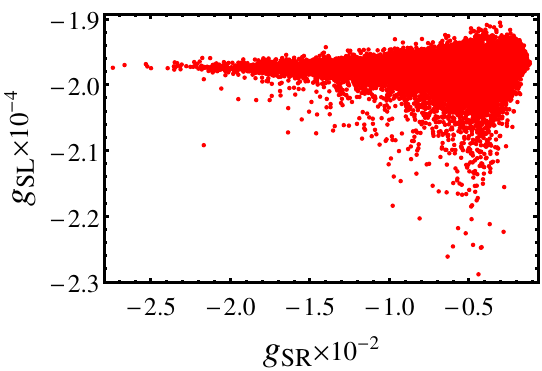}
	\vspace{-0.2cm}
	\caption{(Left) The correlation between the SUSY corrected values of  $g_{VL}$ and $g_{VR}$ is displayed in the right panel, where yellow points represent the configurations that yield  $\Gamma(\tau\rightarrow\mu\nu_{\tau}\nu_{\mu})/\Gamma(\tau\rightarrow e\nu_{\tau}\nu_{e})$ within experimental limits while the red ones are the complete sample. (Right) The correlation between the SUSY corrected values of  $g_{SL}$ and $g_{SR}$ is displayed in the right panel.}
	\label{fig6}
\end{figure}
\begin{figure}[!t]
	\centering
	\includegraphics[height=5cm,width=14.5cm]
	{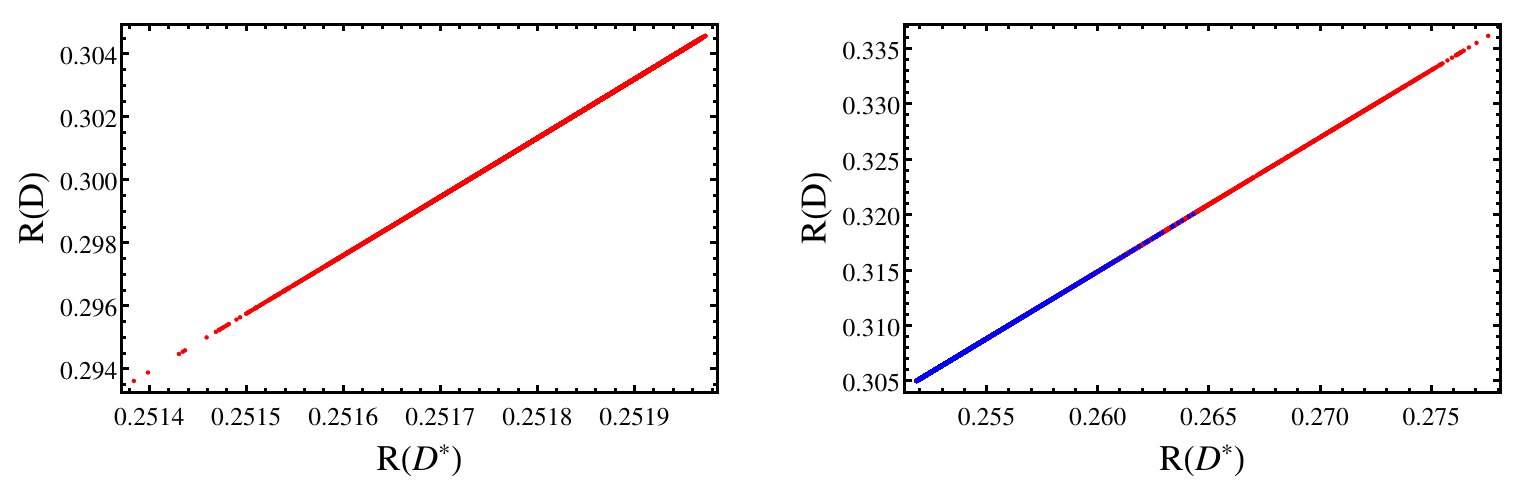}\\	
	\includegraphics[height=5cm,width=15.5cm]
	{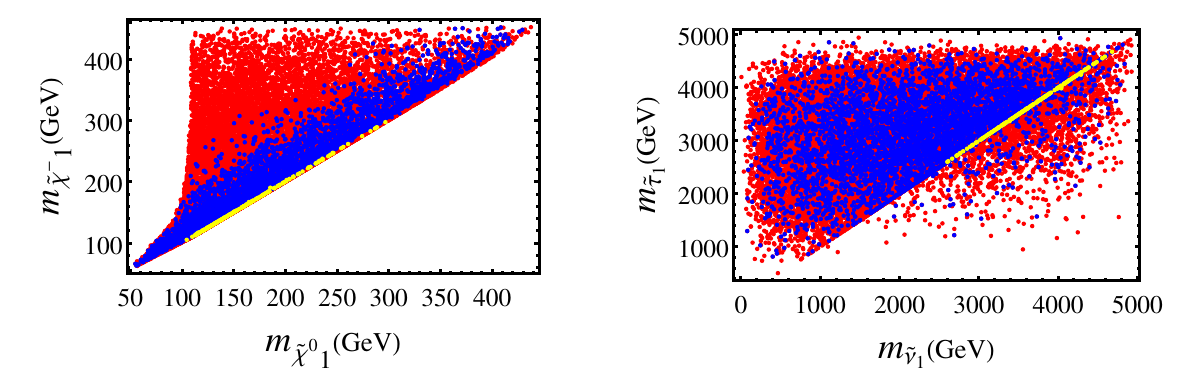}
	\vspace{-0.3cm}
	\caption{The correlation between ${R}(D)$ and ${R}({D^*})$ at tree level (top-left) and after the one-loop
		SUSY contributions through the lepton penguins (top-right) {where the blue points show the constrained ones by  $\Gamma(\tau\rightarrow\mu\nu_{\tau}\nu_{\mu})/\Gamma(\tau\rightarrow e\nu_{\tau}\nu_{e})$. {The correlation between sneutrino and stau masses is on the bottom-left (same colour scheme as in the top-right frame with the additional yellow points representing the region with ${R}({D})>0.33$). (Bottom-right) This represents the correlation between chargino and neutralino masses where the colours have the same meaning as in the bottom-left panel.}}}
	\label{fig4new}
\end{figure}

In essence, the one-loop SUSY effects onto the $W^\pm$ widths are scaled by the $W^\pm$ squared mass while in ${R}(D)$ and ${R}(D^*)$ only by the meson squared masses.
These suppressions are crucial for satisfying the experimental constraints on the ratio of the $W^\pm$ decay widths so that the results of ${R}(D)$ and ${R}({D^*})$ can be accommodated in unexcluded regions of the MSSM parameter space.

As mentioned, the enhancement of $C_{VL}^{\tilde{\tau}}$ occurs mostly when the chargino and neutralino masses are light and similar, in addition to large $\tan\beta$ and stau mass. Therefore, in our scan, we focus on benchmark points where the gaugino soft masses {are given  by $M_1$, $M_2$ $\in [110,500]$ GeV and $M_3 =2$ TeV. Also, we choose the $\mu$ parameter $\in[100,500]$ GeV, $m_{A^{0}}^2 \in [0,25\times10^{4}]$ GeV${^2}$, the $A$ terms $\in [-2000,-100]$ GeV, $M_{\tilde{Q}}$, $M_{\tilde{U}}$ and $M_{\tilde{D}}$ are fixed in the TeV range
	while the slepton soft mass terms $m_{\tilde{L}}$ and $m_{\tilde{E}} \in [100, 5000]$ GeV. Finally, we take $\tan \beta \in [5, 70]$. (As mentioned, the aforementioned Tab.~\ref{tab2} shows an example yielding large corrections to our two observables
	extracted from such a scan.)}

In Figs.~\ref{fig4} and \ref{fig5} we display the regions in the  ($g_{SL}, g_{SR}$)  and ($g_{VL}, g_{VR}$) planes, respectively, that can accommodate the BaBar, Belle and combined average results on ${R}(D)$ and ${R}({D^*})$ within a $1 \sigma$ and $2\sigma$ CL. We also compare these ranges with the MSSM expectations at the one-loop level.  It is clear that the contributions that induce vector operators, like the aforementioned triangle	
 diagrams, lead to ${R}(D)$ and ${R}({D^*})$ close to or potentially within the experimental regions. We can also conclude that $g_{VL}$ must be non-vanishing and of order $0.1$ while $g_{VR}$ can be in the range $[-0.1,0.1]$. This conclusion is explicitly confirmed in Fig.~\ref{fig6}, where the correlation between the SUSY contribution to $g_{VL}$ and $g_{VR}$ is presented. As expected, $g_{VR}\sim 0$ in the MSSM, which has no right-handed vector contribution, while $g_{VL}$ can be of order few percents, which can account for the Belle results within the $1\sigma$ limit and on the 
borderline with the $2\sigma$ band of BaBar and average results. Herein, SUSY contributions to $g_{SL}$ and $g_{SR}$, which are negligibly small, $\sim 10^{-4}$, are also displayed.
 

In the top-left (top-right) panel of Fig.~\ref{fig4new} we present the correlation between ${R}(D)$ and ${R}(D^*)$ at tree-level (at one-loop due to the SUSY contributions to the lepton penguins alone). As  can be seen from this plot, in presence of MSSM one-loop corrections,
	${R}(D)$ can reach $0.335$ while ${R}(D^*)$ extends to $0.277$, which are results
	rather consistent with the Belle measurements and not that far from the BaBar ones. {This correlation can be understood from the fact that SUSY one-loop corrections give a significant contribution to $g_{VL}$ only (of order $6\%$) and, hence, according to Eqs. (15)--(16), both $R(D)$ and $R(D^*)$ are affected by the same correction factor $\propto (1+g_{VL})^2$ through a common Wilson coefficient.
	It is also worth noting that the enhancements of $R(D)$ and $R(D^*)$ require a very peculiar region of parameter space of the MSSM, especially in terms of
$m_{{\tilde\chi}_1^-}$ and $\tan\beta$, wherein, however, all experimental and theoretical constraints sensitive to the latter two quantities are taken into account and included in our scan and numerical analysis.}	
	To our knowledge, these enhancements in both  ${R}(D)$ and ${R}(D^*)$ have never been accounted for before in any NP scenario. The dependence of ${R}(D)$ and ${R}(D^*)$ on $\tan \beta$ is displayed in Fig. \ref{fig:tanbeta}. As can be seen from these plots, a larger value of $\tan \beta$ is preferred by larger values of ${R}(D)$ and ${R}(D^*)$. This can be understood from Eqs. (\ref{chargino_coup}) and (\ref{neutralino_coup})  that emphasise the increase of the neutralino and chargino couplings with the $\tau$ lepton at very large $\tan \beta$.

\begin{figure}[!h]
	\centering
	\includegraphics[height=5.5cm,width=7.5cm]
	{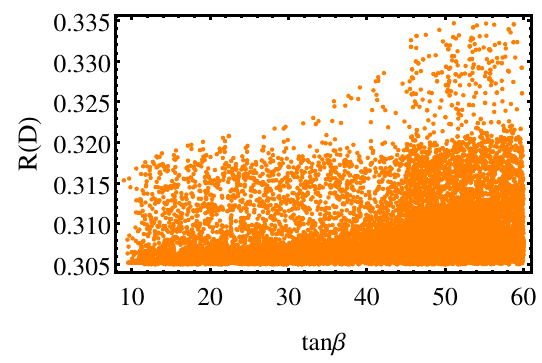}~~~~	
	\includegraphics[height=5.5cm,width=7.5cm]
	{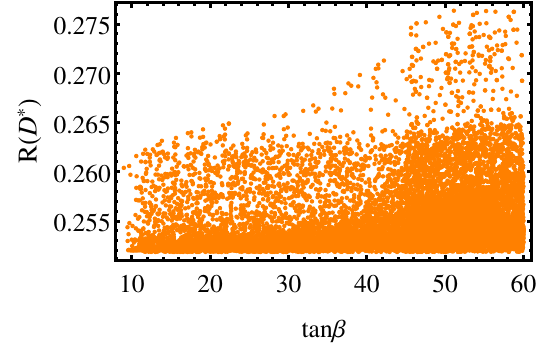}\\
\vspace{-0.3cm}
	\caption{${R}(D)$ and ${R}({D^*})$ as functions of $\tan \beta$. The scan over the parameter space is performed as in the previous plots.}
	\label{fig:tanbeta}
\end{figure}

It is also very  relevant to extract the typical mass spectra which are responsible for the MSSM configurations yielding
${R}(D)$ and ${R}(D^*)$ values (potentially)
consistent with experimental measurements, as these might be accessible during Run 3 at the LHC. As an indication, this is done in Fig.~\ref{fig4new} (bottom-left panel) for the case of the lightest
stau and sneutrino.
The plot shows a predilection of the highest ${R}(D)$ and ${R}(D^*)$ points for MSSM parameter configurations with $m_{\tilde\tau_1}>m_{\tilde\nu_1}$ while the absolute mass scale can cover the entire interval from 200 GeV to 5 TeV. However, the points with ${R}(D)>0.33$ require a rather large $\tilde\tau_1$ mass (say above 2.5 TeV) irrespectively of the  $\tilde\nu_1$ one as well as large $\tan\beta$. This signals that there occurs an interplay between mass suppressions in the loops and enhancements in the couplings.  

\section{Conclusions}\label{sec6}

In summary, we have proven that the MSSM has the potential to alleviate the anomaly presented by  recent data produced by especially  Belle and (somewhat less so) BaBar, which revealed a rather significant
excess above and beyond the best SM predictions available in the observed {\rm BR}($\bar{B}\to D \tau\bar{\nu}_{\tau}$) and {\rm BR}($\bar{B}\to D^{*} \tau\bar{\nu}_{\tau}$) relative to the light lepton cases.
Most remarkably, within the MSSM, the excesses can be explained {\sl simultaneously}, needless to say, over the same regions of  parameter space. Further, the latter do not correspond to any particularly
fine-tuned dynamics {(possibly apart from light neutralino/chargino masses, plus a preference for heavy $\tilde\tau_1$ and $\tilde\nu_{1}$, recall Fig.~\ref{fig4new})}
and a more than acceptable agreement with the Belle (especially) and BaBar (to a lesser extent)  data can be reached via MSSM spectra easily compatible with current experimental constraints from a variety of sources (flavour physics, Higgs boson measurements, SUSY searches). Such a conclusion is obtained after the {\sl first complete} tree-level plus (penguin dominated) one-loop calculation of all MSSM topologies entering the partonic  $b \to c \tau \nu_\tau$ decay process matched with standard computational elements enabling the transition from the partonic to hadronic level.  If forthcoming  data will confirm the anomalous BaBar and Belle results, {e.g.}, from the now running LHCb experiment at the LHC, our findings are rather interesting since a variety of other (typically non-SUSY) models have been tried and tested as a possible explanation of the $\bar{B}\to D \tau\bar{\nu}_{\tau}$ and $\bar{B}\to D^{*} \tau\bar{\nu}_{\tau}$ anomalies and failed.
On the one hand, our results might then be taken as a circumstancial evidence of SUSY. On the other hand, they might pave the way to its direct discovery as
they point to spectra in the sparticle sector of the MSSM that can be accessed at Run 3.

\section*{Acknowledgments}

SK is supported from the STDF project 13858. DB is supported by the Algerian Ministry of Higher Education and Scientific Research under the PNE Fellowship and CNEPRU Project No. B00L02UN180120140040. SM is supported through the NExT Institute. All authors acknowledge support
from the grant H2020-MSCA-RISE-2014 n. 645722 (NonMinimalHiggs) and thank A. Vicente and F. Staub for help.

\begin{table}[!t]
	\centering
	\begin{tabular}{|c|c|c|}
		\hline
		{Loop function}    &{$b \to c \tau \nu_{\tau}$} 	&{$W \to \tau \nu_{\tau}$}\\[0.1cm]
		\hline $C_{00}$ & $-1.2880$ & $-1.2879
		$\\[0.1cm]
		$C_{0}$ & $2.910\times 10^{-7}~\mathrm{GeV^{-2}}$ & $2.931\times 10^{-7}~\mathrm{GeV^{-2}}$\\[0.1cm]
		$B_{0}$ & $0.0281$ & $0.0715$\\
		\hline
	\end{tabular}
	\caption{Loop functions values which correspond to the benchmark given in  Tab.~\ref{tab2}, {where the renormalisation scale, $\widetilde \mu$, has been fixed at 1 GeV}.}
	\label{tab1}
\end{table}
\begin{table}[H]
	\centering
	\begin{tabular}{|c|c|}
		\hline
		{Parameter} &{Value} \\[0.1cm]
		\hline
		\hline
		$\tan\beta$  & $55.22$ \\[0.1cm]
		$\mu$ &  $432.05~{\rm GeV}$\\[0.1cm]
		$M_{A^{0}}^2~({\rm tree})$ & $2.5\times 10^5~{\rm GeV}$\\[0.1cm]
		$M_{1},~M_{2},~M_{3}$ &$326.59,~169.40,~2000~{\rm GeV}$\\[0.1cm]
		$M_{\widetilde{U}},~M_{\widetilde{D}},~M_{\widetilde
			{Q}}$ & $1~{\rm TeV}$ (all)\\[0.1cm]
		$ M_{\widetilde{E}_{1}},~M_{\widetilde{E}_{2}}
		,~M_{\widetilde{E}_{3}}$ &$1693.70,~4926.31,~1033.35~{\rm GeV}$\\[0.1cm]
		$M_{\widetilde{L}_{1}},~M_{\widetilde{L}_{2}}
		,~M_{\widetilde{L}_{3}}$ & $4877.02,~4910.96,~4446.64~{\rm GeV}$\\[0.1cm]
		$A_{t},~A_{b},~A_{\tau}$ & $-1806.83,~-1887.99,~0~{\rm GeV}$\\[0.1cm]
		\hline $m_{\widetilde{\chi}_{1}^{0}},~m_{\widetilde{\chi}
			_{1}^{-}}$ &  $173.07,~173.52~{\rm GeV}$\\[0.1cm]
		$m_{\widetilde{\nu}_{1}},~m_{\widetilde{\tau}_{1}}$  & $4466.25,~4467.40~{\rm GeV}$\\[0.1cm]
		$M_{h^{0}},~M_{H^{0}},~M_{A^{0}},M_{H^{-}}$ &
		$125.13,~441.72,~441.87,~405.60~{\rm GeV}$\\[0.2cm]
		\hline
		$\Gamma_{\tilde{\chi}_{1}^{-}\nu_{\tau}\tilde{\tau}^{\ast}_{1}}^{L}$ & 0.629 \\[0.3cm] $\Gamma_{\tilde{\chi}_{1}^{-}\bar{\tau}\tilde{\nu}_{1}}^{R}$ &-0.656 \\[0.3cm]
		$\Gamma_{\bar{\tau}\tilde{\chi}_{1}^{0}\tilde{\tau}_{1}}^{R}$& -0.447\\[0.3cm] $\Gamma_{\nu_{\tau}\tilde{\chi}_{1}^{0}\tilde{\nu}_{1}^{\ast}}^{L}$&-0.460\\[0.3cm] $\Gamma_{\tilde{\chi}_{1}^{+}\chi_{1}^{0}W^{-}}^{R}$&-0.642\\[0.3cm] $\Gamma_{\tilde{\chi}_{1}^{+}\chi_{1}^{0}W^{-}}^{L}$&-0.642\\[0.3cm]
		$\Gamma_{\bar c bW^{-}}^{L}$&-0.019\\[0.1cm]
		\hline
		\
		$ g_{VL}^{\tilde{\tau}}(b\rightarrow c\tau\overline{\nu}_{\tau})$ &
		$0.021$\\[0.1cm]
		$ g_{VL}^{\tilde{\nu}}(b\rightarrow c\tau\overline{\nu}_{\tau})$ &
		$0.022$\\[0.1cm]
		$ g_{VL}^{H^0,A^0,H^-}(b\rightarrow c\tau\overline{\nu}_{\tau})$ &
		$0.0042$\\[0.1cm]
		\hline
		$g_{VL}(b\rightarrow c\tau\overline{\nu}_{\tau})=g_{VL}^{\tilde{\tau}}+g_{VL}^{\tilde{\nu}}+g_{VL}^{H^0,A^0,H^-}$  & $0.047$\\[0.15cm]
		$g_{VL}(b\rightarrow ce\overline{\nu}_{e})$  & $0$\\[0.15cm]
		\hline
		$ {R}(D)$ & $ 0.335$\\[0.1cm]
		${R}(D^{\ast})$& $0.276$\\[0.1cm]
		\hline
	\end{tabular}
	\caption{Illustrative benchmark point yielding large $ {R}(D)$ and ${R}(D^{\ast})$ values.}
	\label{tab2}
\end{table}


\end{document}